\RequirePackage{fix-cm}
\documentclass[english]{article}
\usepackage[T1]{fontenc}
\usepackage[latin9]{inputenc}
\usepackage{babel}
\usepackage{textcomp}
\usepackage{endnotes}
\usepackage{amstext}
\usepackage{amssymb}
\usepackage{graphicx}
\usepackage[unicode=true,pdfusetitle,
 bookmarks=true,bookmarksnumbered=false,bookmarksopen=false,
 breaklinks=false,pdfborder={0 0 1},backref=false,colorlinks=false]
 {hyperref}
\usepackage{breakurl}
\usepackage{cite}
\usepackage[]{subfig}

\makeatletter

\newcommand{\lyxmathsym}[1]{\ifmmode\begingroup\def\b@ld{bold}
  \text{\ifx\math@version\b@ld\bfseries\fi#1}\endgroup\else#1\fi}

 \let\footnote=\endnote
\newcommand{\lyxaddress}[1]{
\par {\raggedright #1
\vspace{1.4em}
\noindent\par}
}

\usepackage{lmodern}
\usepackage{fancyhdr}

\makeatother

\begin{document}

\title{Water activity in lamellar stacks of lipid bilayers:
  ``Hydration forces'' revisited}

\author{R.~Leite Rubim$^{1,2}$, B.B.~Gerbelli$^1$, K.~Bougis$^{1,2}$,
  C.L.~Pinto de Oliveira$^1$, \\ L.~Navailles$^2$, F.~Nallet$^2$ and
  E.~Andreoli de Oliveira$^1$}

\maketitle

\begin{abstract}
  Water activity and its relationship with interactions stabilising
  lamellar stacks of mixed lipid bilayers in their fluid state are
  investigated by means of osmotic pressure measurements coupled with
  small-angle x-ray scattering. The (electrically-neutral) bilayers
  are composed of a mixture in various proportions of lecithin, a
  zwitterionic phospholipid, and Simulsol, a non-ionic cosurfactant
  with an ethoxylated polar head. For highly dehydrated samples the
  osmotic pressure profile always exhibits the ``classical''
  exponential decay as hydration increases but, depending on Simulsol
  to lecithin ratio, it becomes either of the ``bound'' or ``unbound''
  types for more water-swollen systems. A simple thermodynamic model
  is used for interpreting the results \emph{without} resorting to the
  celebrated but elusive ``hydration forces''.
\end{abstract}

\lyxaddress{$^1$Universidade de S\~ao Paulo, Instituto de F\'\i
  sica-GFCx, P.O.B.  66318, S\~ao Paulo, SP 05314-970, Brazil.}

\lyxaddress{$^2$Universit\'e de Bordeaux, Centre de recherche
  Paul-Pascal\textendash{}CNRS, 115 avenue du Docteur-Schweitzer,
  F-33600 Pessac, France.}

\section{Introduction }
\label{intro}
From mixtures of lipids and water, multilamellar systems may naturally
emerge. In such self-assembled systems, a periodic structure is formed
by stacking lipid bilayers and layers of water. In the case where
there is no in-plane order in the bilayers--in their so-called ``fluid
state''--, the system exhibits the symmetry of a smectic~A phase,
commonly referred to as a lamellar L$_{\alpha}$ phase in the context
of lipid materials~\cite{gulik1967}.
\par Assessing the mechanisms responsible for the stability of such
lamellar structures has been the core motivation for quite a large
number of experimental or theoretical studies, with ``hydration
interactions''~\cite{leNeveu1977}, on one hand, and ``undulation
interactions''~\cite{helfrich1978} on the other hand emerging as
central concepts more than 35 years ago when direct electrostatic
interactions are irrelevant. Techniques of choice for studying
inter-bilayer interactions include osmotic pressure
control~\cite{leNeveu1977,leNeveu1976,parsegian1979}, line-shape
analysis in high-resolution~\cite{safinya1986} or grazing-incidence
small-angle scattering geometries~\cite{salditt2002}, dynamic light
scattering~\cite{nallet1989}, and ``direct'' methods with, \emph{e.g.}
Surface Force Apparatuses~\cite{richetti1990}. Clear experimental
evidences have been obtained as regards undulation interactions in the
swollen end of dilution lines~\cite{safinya1986}, while the other
(highly dehydrated) limit has often been generically characterised in
terms of ``hydration interactions'' from the observed
exponentially-decaying force (or pressure) profile with natural scale
0.2--0.4~nm~\cite{leNeveu1977}.
\par Here, we study a lecithin-based lamellar system in the presence
of a non-ionic co-surfactant, a system which has shown its ability to
efficiently encapsulate DNA fragments in spite of the absence of any
obvious direct electrostatic mechanism at
play~\cite{pott2003,oliveira2010,teixeira2011}. The co-surfactant we
use is an ethoxylated fatty acid, \emph{i.e.} a (short) non-ionic
block copolymer with amphiphilic properties. The present work
therefore somehow expands Ref.~\cite{gerbelli2013,bougis2015}, the
focus being now \emph{osmotic pressure control} instead of line-shape
analysis in small-angle scattering.
\par The lamellar structure of the stacked bilayers is equilibrated
with various semi-dilute (aqueous) solutions of polymers, which gives
a handle on water activity (or, equivalently, osmotic pressure $\Pi$),
while the stacking period $\ell$ is determined by means of small-angle
X-ray scattering, following the method popularised by
V.A.~Parsegian~\cite{parsegian1979}. Other relevant quantities are
commonly manipulated when presenting or discussing the results, namely
the bilayer volume fraction $\phi$ and the interfacial area per
(average) amphiphilic molecule $\bar{\Sigma}$. Assuming homogeneous
and ideally flat bilayers, a simple geometric description of the
lecithin--Simulsol lamellar stack in water gives (see, \emph{e.g.},
Ref.~\cite{bougis2015})
\begin{equation}
  \ell=2\frac{\bar{v}}{\bar{\Sigma}}\times\frac{1}{\phi}
  \label{eq:geom}
\end{equation}
where $\bar{v}$ is the effective molecular volume $xv_S+(1-x)v_L$ of
the bilayer species (with $v_L$, respectively $v_S$, being the
molecular volume of the lecithin, resp. Simulsol molecules and $x$ the
Simulsol mole fraction in the bilayer), $\bar{\Sigma}$ similarly being
an effective interfacial molecular area $x\Sigma_S+(1-x)\Sigma_L$
derived from actual Simulsol $\Sigma_S$ and lecithin $\Sigma_L$
interfacial areas. Note that, while $\Pi$, $\ell$, $x$ or $\phi$ are
experimentally well-defined quantities, and $v_L$, as well as $v_S$,
can safely be considered as \emph{constant} parameters, $\bar{\Sigma}$
is \emph{model-dependent}, as should be clear from the assumptions
leading to eq.~(\ref{eq:geom}): These assumptions would be spoilt by
the presence of numerous structural defects (holes across single
bilayers, passages connecting adjacent bilayers--implying no longer
homogeneous bilayers), or large-amplitude area-storing bilayer
undulations--implying no longer flat bilayers.
\section{Thermodynamic considerations}
\label{thermo}
\subsection{Lamellar stacks}
\label{lam-stacks}
Implicitly assuming for the sake of simplicity a two-component
(lipid--water) system, it is customary to cast the interpretation
framework of the $\Pi(\ell)$ data into the mould of, loosely speaking,
an inter-bilayer interaction potential energy per unit bilayer area
$V$, but perhaps more rigorous (as already noticed in
Ref.~\cite{parsegian1979}) to start from the excess \emph{free energy}
of the bilayer stack. Per unit volume of the lamellar stack (and
disregarding an obvious dependence on temperature that remains
implicit in the following), the excess free energy is a function of
two among the three quantities appearing in
eq.~(\ref{eq:geom}). Choosing $\phi$ and $\bar{\Sigma}$, the excess
free energy density, formally written as follows:
\begin{equation}
f_{\mathrm{exc}}(\phi,\bar{\Sigma})
\label{eq:fexces}
\end{equation}
should be zero, by definition, for either $\phi=0$ or $\phi=1$. The
total Gibbs free energy is then expressed as
\begin{equation}
  G=(N_wv_w+N_l\bar{v})\left[p+\frac{\mu_w^+}{v_w}(1-\phi)+\frac{\mu_l^+}{\bar{v}}\phi+f_{\mathrm{exc}}(\phi,\bar{\Sigma})\right]
\label{eq:gibbs}
\end{equation}
Here, $p$ is the (hydrostatic) pressure. Indices $w$, respectively
$l$, refer to water, resp. ``average'' lecithin / Simulsol lipid. The
total numbers of molecules of a given species are $N_w$ and $N_l$, and
the chemical potentials for the \emph{pure} species are $\mu_w^+$ and
$\mu_l^+$. The lipid volume fraction is
$\phi=N_l\bar{v}/(N_wv_w+N_l\bar{v})$. Incompressibility is enforced
by considering the molecular volumes $v_w$ and $\bar{v}$ as constant
parameters, with therefore $\partial G/\partial p=N_wv_w+N_l\bar{v}$.
\par The ``dilution law'', namely the $\ell(\phi)$ relation, is
derived from eq.~(\ref{eq:geom}) once the optimal interfacial area is
obtained, for a given composition $\phi$, by solving the minimisation
equation
\begin{equation}
\frac{\partial G}{\partial \bar{\Sigma}}=0
\label{eq:optSigma}
\end{equation}
which is equivalent to solving
\begin{equation}
\frac{\partial f_{\mathrm{exc}}}{\partial \bar{\Sigma}}=0
\label{eq:optSigma2}
\end{equation}
\par The ``bilayer equation of state'', that is to say the explicit
solution to this latter equation, is commonly expressed as
$\bar{\Sigma}(\phi)$ or $\bar{\Sigma}(\ell)$ relations.
\par The ``lamellar stack equation of state'' results from equating
the water chemical potential in a lamellar stack of given composition
$\phi$--and, therefore, of given optimal interfacial area
$\bar{\Sigma}(\phi)$--submitted to the (hydrostatic) pressure $p+\Pi$
to the pure water chemical potential when pure water is submitted to
(hydrostatic) pressure $p$. The resulting standard expression is
\begin{equation}
  \Pi=\phi\frac{\partial f_{\mathrm{exc}}}{\partial
    \phi}-f_{\mathrm{exc}}
  \label{eq:posm}
\end{equation}
either expressed as a $\Pi(\phi)$ or a $\Pi(\ell)$ relation.
\par
\begin{figure}[h]
\begin{centering}
  \subfloat[][Osmotic pressure
  $\Pi$]{\includegraphics[scale=1,width=0.45\columnwidth]{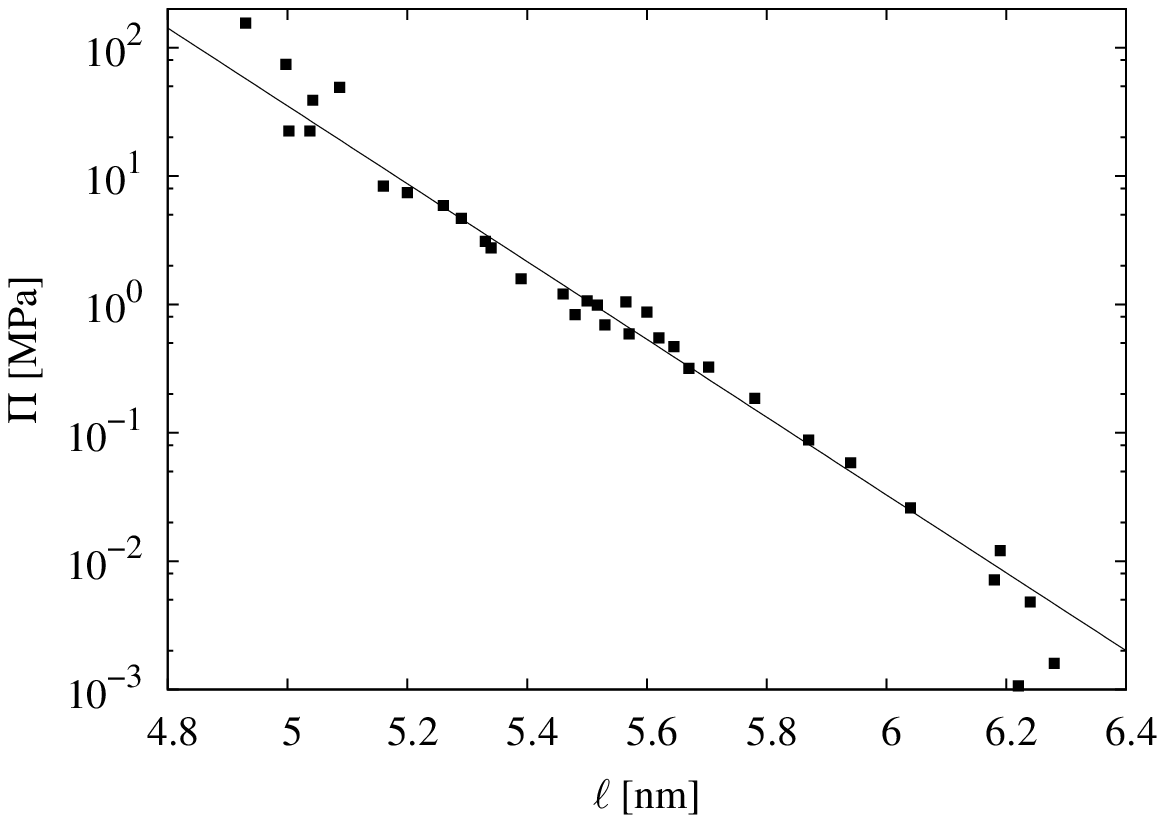}\label{fig:parsegianPi}}
  \qquad\subfloat[][Lipid interfacial area
  $\bar{\Sigma}$]{\includegraphics[scale=1,width=0.45\columnwidth]{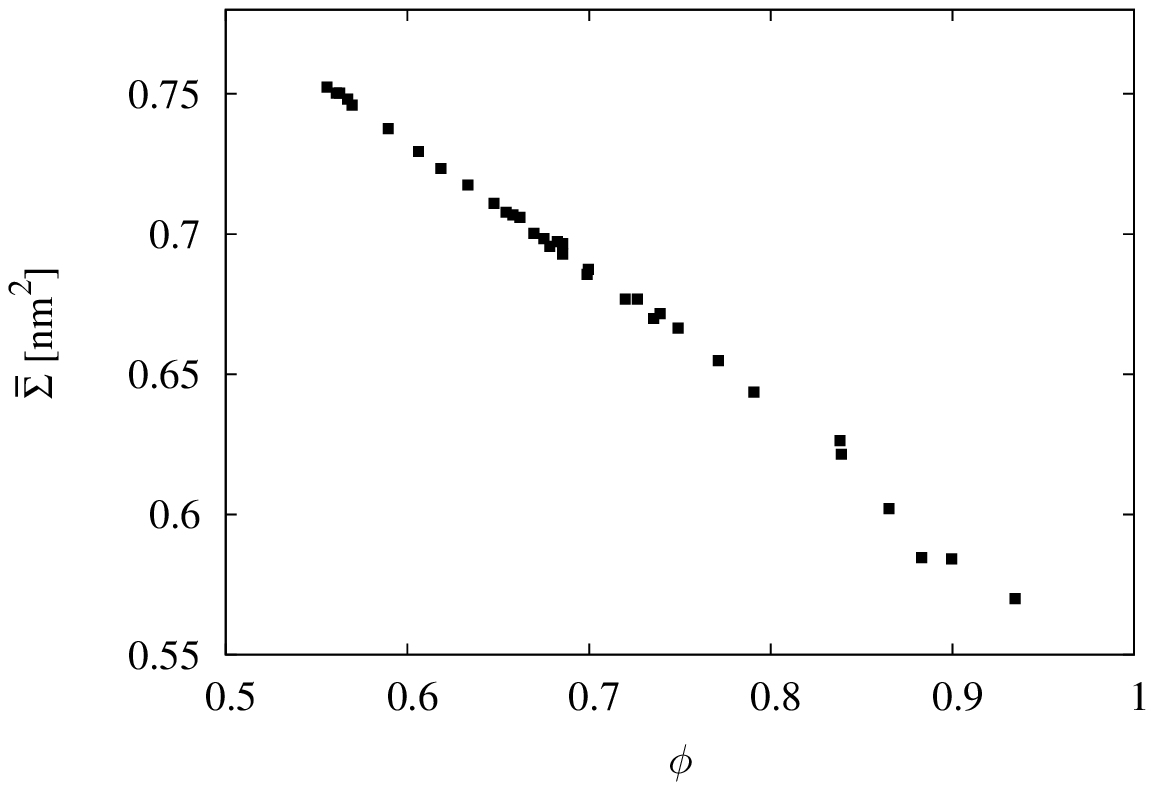}\label{fig:parsegianSigma}}
\end{centering}
\caption{(a) Lamellar stack equation of state. Note the
  quasi-exponential decay of osmotic pressure with stacking period
  $\ell$--continuous line. Decay length \emph{ca.} 0.143~nm. (b)
  Bilayer equation of state. Dehydration results in a \emph{decrease}
  in optimal interfacial area per lipid molecule. Redrawn from data
  for egg lecithin in Ref.~\cite{parsegian1979}\label{fig:parsegian}}
\end{figure}

 Examples of the two above-mentioned equations of state are given
for illustration in Fig.~\ref{fig:parsegianPi} and
\ref{fig:parsegianSigma}--data extracted from
Ref.~\cite{parsegian1979}.
The quasi-exponential decay of the osmotic pressure $\Pi$ with
stacking period $\ell$ is also observed when $\Pi$ is plotted against
the thickness of the water channel $\ell_w\equiv\ell(1-\phi)$, though
with a different decay length: A value \emph{ca.}  0.256~nm is found
in the latter case~\cite{parsegian1979}, instead of \emph{ca.}
0.143~nm in Fig.~\ref{fig:parsegianPi}. Quite significant also is the
\emph{decrease} in optimal interfacial area per lipid molecule
$\bar{\Sigma}$ as dehydration proceeds, a feature already observed
more than 50 years ago with surfactant--water or lipid--water
systems~\cite{luzzati1960,luzzati1962,reiss-husson1967}.
\par In the quite common case where a phase coexistence between a
lamellar structure with finite spacing $\ell_{\mathrm{max}}$ and
(almost) pure water when added in excess is observed, the lamellar
stack is said to be ``bound''. This amounts to saying that, at some
``dilution limit'', there is a finite value $\phi^*$ for the lipid
volume fraction where the osmotic pressure reaches zero:
$\Pi(\phi^*)=0$. In the system studied in Ref.~\cite{parsegian1979},
for instance, the dilution limit is found at about
$\ell_{\mathrm{max}}=6.25$~nm. At contrast, (less common) ``unbound''
systems may be swollen seemingly indefinitely with water, and
$\phi^*\rightarrow0$ or, equivalently,
$\ell_{\mathrm{max}}\rightarrow+\infty$. The so-called ``unbinding
transition'', theoretically described in
Ref.~\cite{lipowski1986,podgornik1992,milner1992}, separates ``bound''
systems, with somehow weak undulation interactions, from ``unbound''
systems where stronger undulation interactions overcome attractive,
van der Waals forces. It has been experimentally evidenced in
appropriately chosen lamellar systems, see for instance
Ref.~\cite{bougis2015} for a recent report.
\subsection{Simple fluids}
\label{fluids}
\begin{figure}
\begin{centering}
  \includegraphics[scale=1,width=\columnwidth]{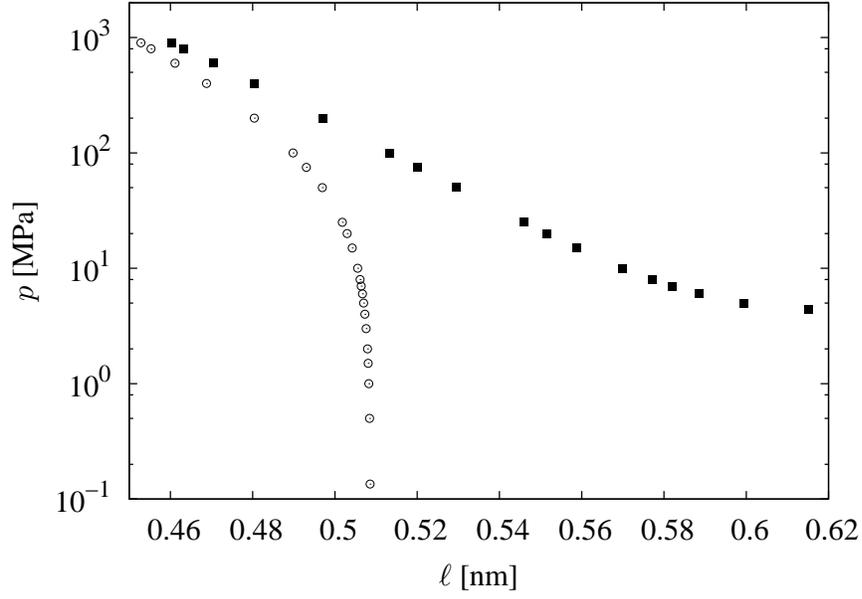}
\end{centering}
\caption{Pressure equations of state for \emph{liquid} ethane
  extracted from Ref.~\cite{buecker2006}, expressed as a function of
  $\ell\equiv\frac{\sqrt{2}}{2}\left(\frac{4M}{\rho
      N_A}\right)^{1/3}$, at two temperatures $T=190$~K ($\odot$) and
  $T=300$~K ($\blacksquare$). For pure ethane, triple point and,
  respectively, critical point temperatures and pressures are
  $T_t=90.4$~K and $p_t=1.14$~Pa, resp. $T_c=305.3$~K and
  $p_c=4.93$~MPa. The $\ell$ variable would be the nearest-neighbour
  distance if the mass density $\rho$ of liquid ethane were obtained
  by placing molecules at the nodes of a face-centred cubic
  lattice. The molar mass of ethane is $M=30.07$~g/mol and $N_A$ is
  the Avogadro constant. Note the quasi-exponential decay of pressure
  for small intermolecular separations--decay lengths \emph{ca.}
  0.020~nm ($T=190$~K) and 0.024~nm ($T=300$~K)\label{fig:ethane}}
\end{figure}

As argued in more details in this Section, the above-described
unbinding features of lamellar stacks upon addition of
  solvent has actually some
similarities with the liquid--gas transition in pure compounds
when hydrostatic pressure is isothermally
decreased: Molecular (translational) kinetic energy (3D
  configurational entropy) then plays the r\^ole of bilayer
undulations in lamellar stacks (1D configurational
  entropy), with cohesive forces--viz. van der Waals
  interactions--acting similarly in both cases. Close to the triple
point temperature, the liquid phase may be said to be
``bound'' in the sense that pressure may be decreased to (almost) zero
while keeping a rather small specific volume for the system. Close to
the critical point temperature, however, the liquid phase
would be characterised as ``unbound'' because the
specific volume may reach large values \emph{even though} a
significant pressure is still applied to the
fluid. Fig.~\ref{fig:ethane} illustrates this very common behaviour in
the case of the fluid phase of ethane--data extracted from
Ref.~\cite{buecker2006}--where, for the sake of comparison with
lamellar phase data, the physically-relevant parameter, namely the
specific volume, has been expressed as an average \emph{distance}
$\ell$ between neighbouring molecules in the liquid.
\par A conspicuous feature of the experimental pressure equations of
state displayed in Fig.~\ref{fig:ethane} may be noted at the
\emph{high} pressure end of the two curves, with a quasi-exponential
behaviour $p(\ell)\propto\exp(-\ell/\Lambda)$ that somehow resembles
the one observed in Fig.~\ref{fig:parsegianPi}. At contrast with
lamellar phase data, however, the characteristic decay length
$\Lambda$ is here \emph{very short}. Values in the range
$0.02$--$0.024$~nm are found for $\Lambda$, the shorter value being
appropriate for the lower temperature.
\par
\begin{figure}
\begin{centering}
  \includegraphics[scale=1,width=\columnwidth]{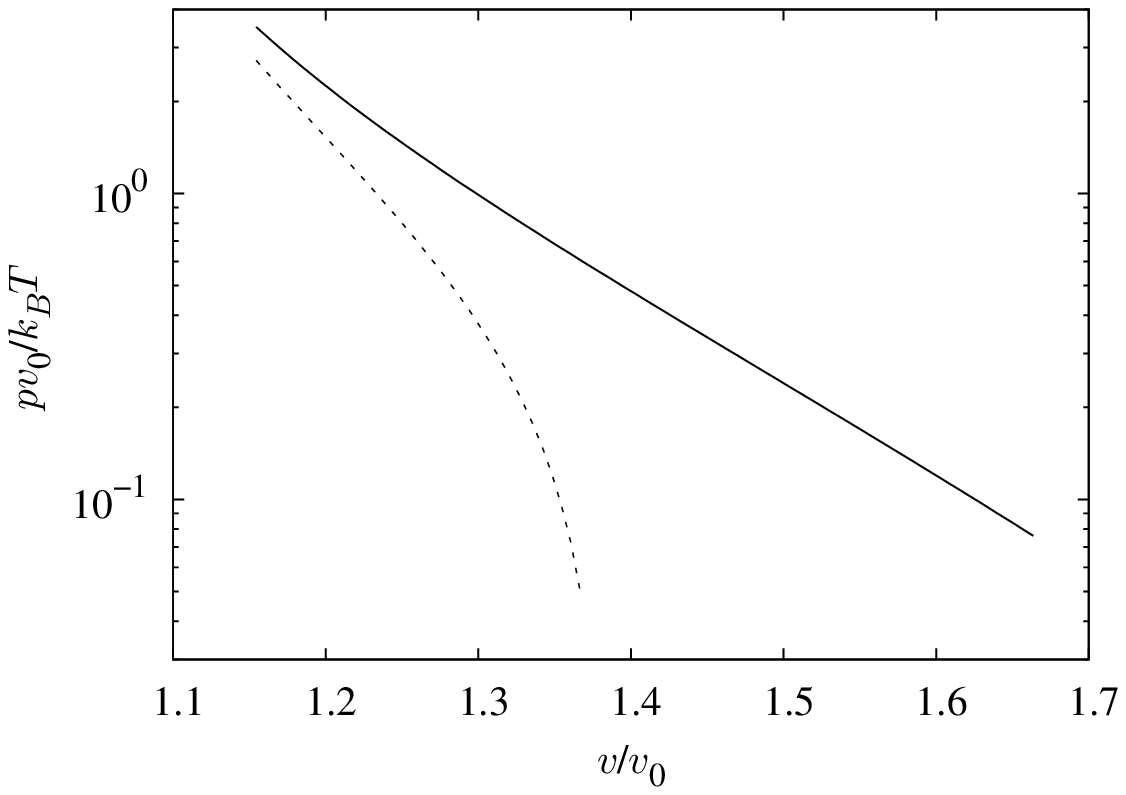}
\end{centering}
\caption{Pressure equations of state in reduced units for the
  \emph{liquid} phase of a van der Waals fluid.  Second virial
  coefficients $b_2/v_0=-10$ (dashed line) and $-7.92$ (continuous
  line). From the common tangent construct, the liquid--gas binodal
  begins, respectively, at the specific volumes $v_L\approx1.37v_0$ and
  $v_L\approx1.67v_0$. Note the quasi-exponential decay of pressure with
  decay parameters, respectively, $0.081v_0$ and $0.143v_0$
  \label{fig:vdw}}
\end{figure}

  Interpreting parameter $\Lambda$ in terms of molecular
  dimensions seems difficult, but a simple interpretation
  for the observed behaviour is easily found in considering the van der Waals
equation of state for pure, fluid compounds. In terms of only two
parameters, namely a second virial coefficient $b_2$ describing the
intermolecular interactions, and an excluded volume $v_0$ accounting
for the finite molecular dimensions, van der Waals equation of state
is~\cite{maxwell1874}
\begin{equation}
  \frac{pv_0}{k_BT}=\frac{1}{2}\frac{b_2}{v_0}\left(\frac{v_0}{v}\right)^2+\frac{v_0/v}{1-\frac{v_0}{v}}
  \label{eq:vdw}
\end{equation}
with $v$ the specific volume of the van der Waals fluid, $k_B$ the
Boltzmann constant and $T$ the temperature. Dense fluids, \emph{i.e.}
systems with specific volumes only moderately larger than $v_0$, may
be obtained at moderate pressures if the second virial coefficient
$b_2$ is \emph{negative} enough. With the integral
  expression for $b_2$ in terms of the Mayer
  function~\cite{mayer1937}, this occurs when
$k_BT$, a measure of thermally-driven translational kinetic
energy, is small enough compared to cohesive energy. Below a critical
temperature $T_c$ where $b_2$ decreases below the condensation
threshold $-27v_0/4$ in the van der Waals model, \emph{liquid} phases
may indeed be formed. An illustration is given in Fig.~\ref{fig:vdw}
for two values of the second virial coefficient, one close to, the
other one farther below the condensation threshold.
Note that, for plotting eq.~(\ref{eq:vdw}) in Fig.~\ref{fig:vdw}, the
specific volume range has been restricted: On the low density end, to
avoid entering the liquid--gas coexistence region where
eq.~(\ref{eq:vdw}) no longer applies, but also on the high density
end, to avoid a spurious logarithmic singularity implicit in the van
der Waals description of the configurational entropy.
\par The similarity between Fig.~\ref{fig:ethane} and \ref{fig:vdw} is
striking, in particular as regards the large mismatch between
  decay parameters, resulting from either experiment or model, and
  characteristic molecular dimensions. A compelling interpretation
for the quasi-exponential pressure decay observed over a range of
intermolecular separations is therefore the following: It is
  not related to any structural property of the system and should be
attributed to the standard competition (with a temperature-dependent
outcome, obviously) between \emph{cohesive} energy, favoured in high
density states because of attractive van der Waals interactions
between molecules ($b_2$ term in eq.~(\ref{eq:vdw})), and
\emph{configurational entropy}, reduced in high density states owing
to short-range steric repulsions ($v_0$ term in
  eq.~(\ref{eq:vdw})). Besides, the precise \emph{shape} of the
  interaction profile, \emph{e.g.}  a $1/\ell^{12}$ repulsive well
  combined with a $-1/\ell^{6}$ attractive component in the standard
  Lennard-Jones parametrisation of van der Waals
  interactions~\cite{jones1924}, is irrelevant here since the
  spatial integration of the Mayer function leading to $b_2$ smoothes
  out such details.
\par Taking one step further and considering the similarities between
Fig.~\ref{fig:parsegianPi} with either Fig.~\ref{fig:ethane} or
Fig.~\ref{fig:vdw}, it now seems natural to relate the
quasi-exponential decay in the lamellar stack equation of state to a
similar competition between cohesive energy (arising in lamellar
stacks from van der Waals interactions through the water channels
between lipid bilayers) and configurational entropy, as
repeatedly suggested by J.~Israelachvili and
H.~Wennerstr\"om~\cite{israelachvili1990,israelachvili1992,israelachvili1996}.
This idea--in essence that ``the swelling process is entropy
driven''\cite{sparr2011}--has found a partial support in recent
  numerical simulations~\cite{schneck2012}. It is developed
in Section~\ref{model} below.
\subsection{Lamellar stacks ``\`a la van der Waals''}
\label{model}
Before presenting our model, a word of caution: Nothing prevents, in
principle, some \emph{direct} (repulsive) ``hydration forces'' playing
a (marginal) r\^ole in the detailed characteristics of the lamellar
stack equation of state. Possible physical mechanisms for such direct forces involve
the solvent (water) binding properties with the bilayer
  surfaces~\cite{kanduc2014}, or solvent structural
properties~\cite{marcelja1976,cevc1982} through a mechanism similar to
wetting~\cite{cahn1958} in this latter case: With
\emph{anti-symmetric} boundary conditions at the bilayer--solvent
interface, the order parameter profile of a (non scalar) property
associated to solvent across the water channel, electric polarisation
typically, would lead indeed to an effective \emph{repulsive}
interaction in the lamellar stack. Interactions would nevertheless
remain essentially attractive in the case of symmetric boundary
conditions, as shown by Richetti \emph{et al.} in their study of a
structure-prone (pre-smectic) solvent confined between solid
surfaces~\cite{richetti1996}. In any case, ``hydration forces'' if
they exist in this restricted sense would merely affect the
\emph{magnitude} of the second virial coefficient in the van der Waals
fluid analogy, increasing $|b_2|$ (if attractive--more cohesion) or
decreasing $|b_2|$ (if repulsive--less cohesion), which would only
slightly displace the location of the critical temperature in the
phase diagram.
\par It therefore appears necessary to reconsider carefully the
interpretation of statements similar to
\begin{quote}
  \textsf{Hydration repulsion universally acts between well-solvated
    surfaces in water and balances the van der Waals attraction in the
    nanometre range}
\end{quote}
(quoted from Ref.~\cite{schneck2012}) found in references spanning
more than 30 years, for instance
Ref.~\cite{parsegian1979,lipowski1986,parsegian2011}. The underlying
vision of a \emph{balance} is commonly expressed in
quantitative terms by adding pressures (forces per unit bilayer area),
with as many contributions to the sum as identified interactions. In
the simplest case where only hydration and van der Waals interactions
are considered to be relevant, the total pressure
is~\cite{parsegian1979}
  \begin{equation}
    \label{eq:pressures}
P(\ell)=P_0\exp\left[-(\ell-\delta)/\lambda\right]-\frac{H}{6\pi}\left[\frac{1}{(\ell-\delta)^3}+\frac{1}{(\ell+\delta)^3}-\frac{2}{\ell^3}\right]
  \end{equation}
where the exponentially-decaying, repulsive term represents hydration
forces and the combination of power laws describes, in the simplest
possible way, attractive van der Waals forces between identical,
planar and parallel objects of infinite lateral extension and
thickness $\delta$ [$\equiv2\bar{v}/\bar{\Sigma}$ in
  eq.~(\ref{eq:geom})], separated by a channel of width
$\ell-\delta$. In eq.~(\ref{eq:pressures}), $P_0>0$ is the amplitude
of the hydration pressure, $\lambda$ its decay length and $H>0$ the
Hamaker coefficient.
\par From Ref.~\cite{parsegian1979}, values appropriate for describing
the experimental measurements in egg-lecithin-based lamellar stacks
are: $P_0\approx7.05\times10^8$~Pa, $\lambda\approx0.256$~nm and
$H\approx6.0\times10^{-21}$~J. As already mentioned in
Section~\ref{fluids}, parameter $\lambda$--introduced here--differs
from parameter $\Lambda$ used for describing the lamellar stack
equation of state $\Pi(\ell)$ displayed in
Fig.~\ref{fig:parsegianPi}. This comes from the choice made in
Ref.~\cite{parsegian1979}, the lamellar stack equation of state being
represented by a \emph{different} function $\Pi(\ell-\delta)$. It
differs from $\Pi(\ell)$ because $\delta$ actually depends on
hydration. The ``balance'' between hydration and van der Waals forces
is obtained at a maximum swelling $\ell_{\mathrm{max}}\approx6.3$~nm,
where bilayer thickness is $\delta\approx3.5$~nm.
\par But there is more in thermodynamics than in
mechanics or, in other terms, free energy is more than
  mere potential energy: A complete description of, \emph{e.g.},
equations of state or phase diagrams is not wholly encapsulated in
force models similar to eq.~(\ref{eq:pressures}) that
  ignore entropy, a criticism already explicitly formulated in the
context of the unbinding transition by R.~Lipowski and
S.~Leibler~\cite{lipowski1986} or S.T.~Milner and
D.~Roux~\cite{milner1992}. They showed that,
even though the mechanical (but \emph{ad hoc}) model,
eq.~(\ref{eq:pressures}), accounts extremely well for the measured
osmotic pressure data--Fig.~\ref{fig:parsegianPi}--it utterly fails in
explaining the detailed thermodynamic features of the unbinding
transition. What is missing in the na\"\i ve
mechanical approach is, specifically, a description of the
\emph{bilayer} equation of state--Fig.~\ref{fig:parsegianSigma}--as
well as a physically well-grounded interpretation of the ``hydration
forces''--two aspects recently emphasised in
References~\cite{parsegian2011,bauduin2014} and \cite{donaldson2015},
respectively.
\par In the present approach, we propose to overcome these
shortcomings by modelling the fundamental thermodynamic quantity in
the problem, namely the excess free energy density $f_{\mathrm{exc}}$,
with the following physical ingredients: i) Helfrich undulation
(entropic) interactions between bilayers across the water channels,
ii) van der Waals (and possibly other)
\emph{direct} interactions at the level of a second virial coefficient
description as in Ref.~\cite{milner1992}, and iii) bilayers described
as being, individually, a kind of two-dimensional van der Waals fluid
well \emph{below} its own unbinding transition in the sense given to
it in Section~\ref{fluids}. It is worth mentioning
  again that, owing to the virial approach chosen in ii), the precise
  \emph{shape} of the interaction potential, a repulsive
  exponentially-decaying term added to an attractive power-law
  contribution as would result from eq.~(\ref{eq:pressures}) for
  instance, is irrelevant here, for the same reasons as given in
  Section~\ref{fluids}. In quantitative terms, our
  heuristic approach amounts to writing
\begin{eqnarray}
\ell\times f_{\mathrm{exc}}
&=&\frac{3\pi^2}{128}\frac{(k_BT)^2}{\kappa}\frac{1}{(\ell-\delta)^2}
\nonumber \\
&&-k_BT\chi\phi^2\ell \nonumber \\
&&+\frac{k_BT}{\bar{\Sigma}}\ln\left[\frac{\varsigma^2}{e(\bar{\Sigma}-\Sigma_0)}\right]+\frac{1}{2}k_BT\frac{b_2}{\bar{\Sigma}^2}
\label{eq:vdw1d-2d}
\end{eqnarray}
where $\ell$ depends on $\bar{\Sigma}$ and $\phi$ according to
eq.~(\ref{eq:geom}) and $\delta$ is again a convenient notation for
$2\bar{v}/\bar{\Sigma}$. In eq.~(\ref{eq:vdw1d-2d}), $e$ being the
base of the natural logarithm, $\kappa$ is the bilayer bending modulus
(controlling the amplitude of thermally-driven bilayer
undulations~\cite{helfrich1978}), $\chi$ is the second virial
coefficient accounting for (inter-bilayer) direct interactions
according to S.T.~Milner and D.~Roux~\cite{milner1992}, $\Sigma_0$ is
the ``excluded area'' in the two-dimensional fluid of lipid molecules,
$b_2$ is the second virial coefficient accounting for
\emph{intra-bilayer} interactions between lipid molecules, ``\`a la
van der Waals'', and $\varsigma$ a parameter with length units
analogous to the thermal de Broglie wavelength in ideal (3D) classical
gases~\cite{kittel1980}.
\par
\begin{figure}
\begin{centering}
  \includegraphics[scale=1,width=\columnwidth]{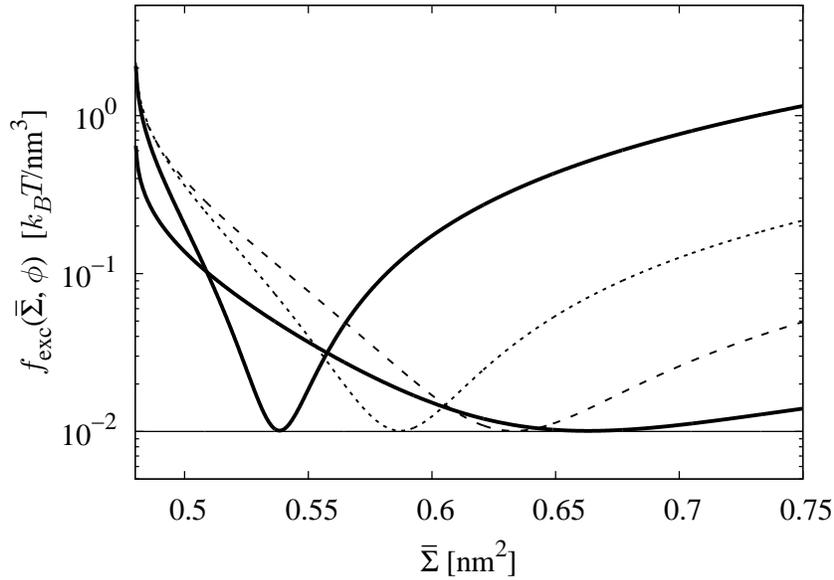}
\end{centering}
\caption{Excess free energy per unit volume of a bilayer stack
  $f_{\mathrm{exc}}$, according to the thermodynamic model
  eq.~(\ref{eq:vdw1d-2d}), drawn as a function of the area
  $\bar{\Sigma}$ per lipid molecule for a few selected values of the
  bilayer volume fraction: $\phi=0.25$--rightmost continuous line,
  $\phi=0.80$--dashed line, $\phi=0.90$--dotted line, and
  $\phi=0.95$--leftmost continuous line. Note that, for the sake of
  clarity, a $\phi$-dependent constant has been added to
  $f_{\mathrm{exc}}$ to adjust the minimum of each curve to the
  (arbitrarily chosen) value 0.01~$k_BT/$nm$^3$ (thin horizontal
  line). The shifts in free energy per unit volume are, along
  decreasing hydration, $\Delta\approx0.46$, 1.43, 1.43 and
  0.87~$k_BT/$nm$^3$
  \label{fig:bilayerOptimalArea}}
\end{figure}

 For illustrating the essential r\^ole of
  ingredient iii) in the physical content of eq.~(\ref{eq:vdw1d-2d}),
Fig.~\ref{fig:bilayerOptimalArea} displays a set of curves for
$f_{\mathrm{exc}}$ as a function of the area $\bar{\Sigma}$ per lipid
molecule for a few selected bilayer volume fractions $\phi$. The
numerical values for the parameters used in the computation of
$f_{\mathrm{exc}}$ are given in Table~\ref{table:params}.
\begin{table}
  \centering
  \begin{tabular}{|c|c|c|c|c|c|} \hline\hline
$\kappa/k_BT$ & $\bar{v}$ & $\Sigma_0$ & $\varsigma$ & $\chi/\varsigma^3$ &
    $b_2/\Sigma_0$ \\ \hline
& [nm$^3$] & [nm$^2$] & [nm] & [nm$^{-6}$] &  \\ \hline\hline
2. & 1.0 & 0.48 & $\sqrt{\Sigma_0}$ & 0.1 & -10. \\ \hline
  \end{tabular}
  \caption{Numerical values chosen for illustrating the properties of
    the thermodynamic model, eq.~(\ref{eq:vdw1d-2d})}
  \label{table:params}
\end{table}
As suggested by the trends observed in
Fig.~\ref{fig:bilayerOptimalArea} (and confirmed by a thorough
numerical analysis), the ``optimal'' area where $f_{\mathrm{exc}}$
reaches its minimum value depends only weakly on the bilayer volume
fraction $\phi$ for \emph{dilute} lamellar stacks, but much more
strongly as the lamellar phase becomes highly dehydrated. This is
easily interpreted, in qualitative terms, by a competition between
\emph{stack} entropy, as described by the Helfrich undulation term
that strives for \emph{large} $\ell$ (corresponding to small
$\bar{\Sigma}$) values, and \emph{two-dimensional} liquid bilayer
entropy, as described by the excluded area $\Sigma_0$ term, oppositely
striving for large $\bar{\Sigma}$ values. The second virial
coefficient $b_2$, negative enough to ensure a ``bound''
two-dimensional liquid state for the lipid fluid, comes into play when
the stack entropy is optimised, \emph{i.e.}
for dilute enough lamellar stacks, and stabilises $\bar{\Sigma}$ to a
value greater than, but still rather close to $\Sigma_0$
($\approx1.4\times\Sigma_0$ for Table~\ref{table:params} parameter
values).
\par
\begin{figure}
\begin{centering}
  \subfloat[][Osmotic pressure
  $\Pi$]{\includegraphics[scale=1,width=0.45\columnwidth]{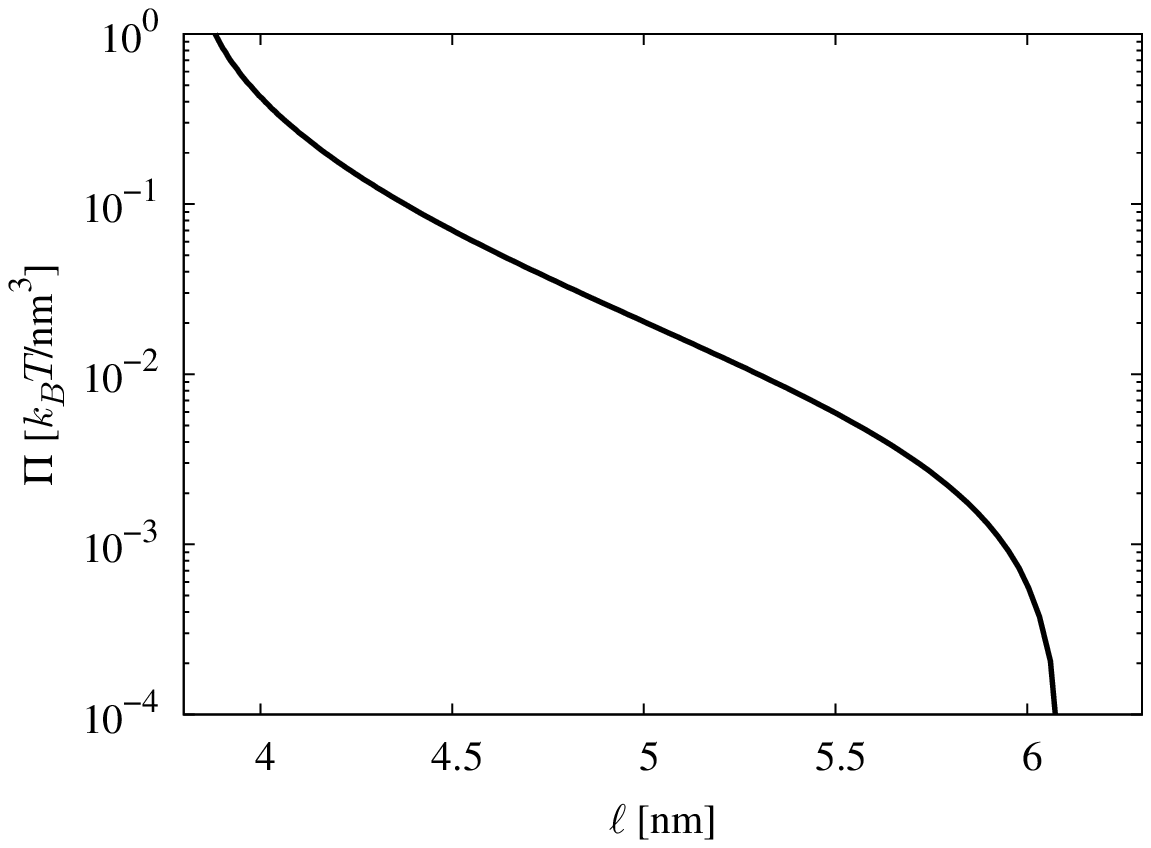}\label{fig:pi}}
  \qquad\subfloat[][Lipid interfacial area
  $\bar{\Sigma}$]{\includegraphics[scale=1,width=0.45\columnwidth]{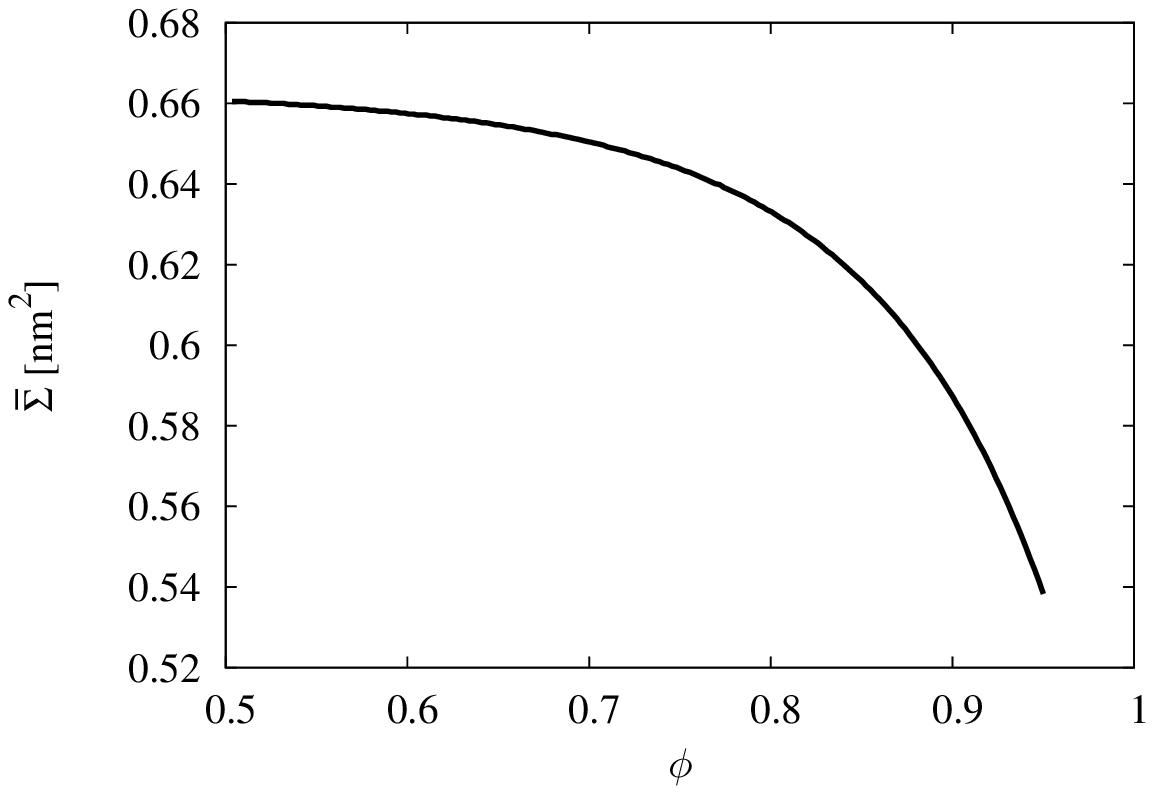}\label{fig:sigma}}
\end{centering}
\caption{(a) Osmotic pressure $\Pi$ of the lamellar stack as a function of
  the stacking period $\ell$. The osmotic pressure reaches \emph{ca.}
  4~MPa when the lamellar phase is highly dehydrated. At the dilution
  limit $\ell_{\mathrm{max}}\approx6.1$~nm, $\Pi\rightarrow0$. There
  is a quasi-exponential decay of the osmotic pressure with a decay
  length $\approx0.43$~nm between these two limits. (b) Optimal area
  $\bar{\Sigma}$ per lipid molecule as a function of the bilayer
  volume fraction $\phi$. Numerical results derived from the
  thermodynamic model eq.~(\ref{eq:vdw1d-2d}) with parameters given in
  Table~\ref{table:params} \label{fig:pi_sigma}}
\end{figure}

\begin{figure}
\begin{centering}
  \includegraphics[scale=1,width=\columnwidth]{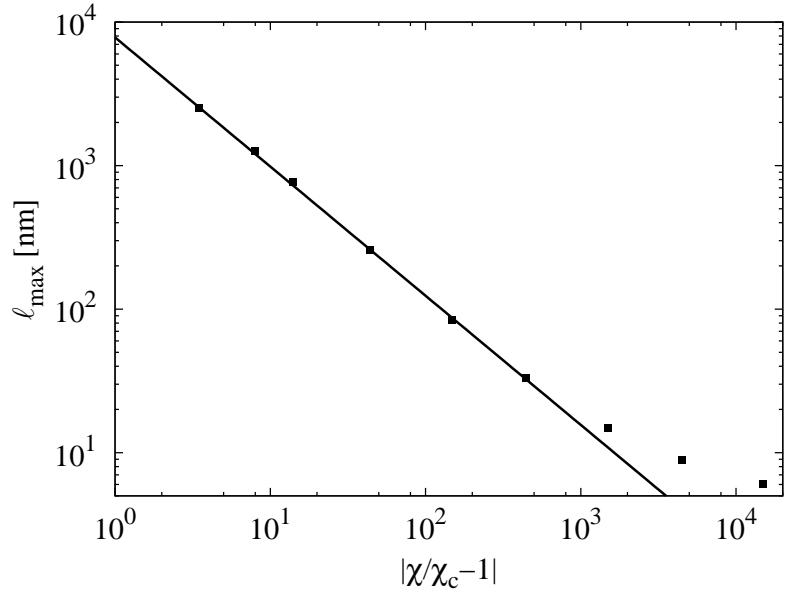}
\end{centering}
\caption{Dilution limit $\ell_{\mathrm{max}}$ as a function of
  $|\chi/\chi_c-1|$. The continuous line describes a power-law
  divergence of $\ell_{\mathrm{max}}$ as the Milner-Roux virial
  coefficient approaches the critical value $\chi_c$, all other
  parameters in eq.~(\ref{eq:vdw1d-2d}) being kept fixed to the values
  given in Table~\ref{table:params}. The power-law divergence, with
  $k_BT\chi_c\approx9.2$~Pa and $\alpha\approx0.9$, is the expected
  signature of a critical unbinding transition occurring in the
  lamellar stack as direct inter-bilayer interactions become less
  attractive~\cite{milner1992}
  \label{fig:unbinding}}
\end{figure}

 Even though it appears difficult to obtain the bilayer equation
of state by analytically solving eq.~(\ref{eq:optSigma2}), a numerical
procedure for determining the optimal $\bar{\Sigma}(\phi)$ is easily
implemented, considering the simple shapes of the $f_{\mathrm{exc}}$
curves displayed in Fig.~\ref{fig:bilayerOptimalArea}. But the
\emph{possibility} of describing a ``bound'' lamellar stack, that is
to say a system where the osmotic pressure $\Pi$ reaches 0 at a
\emph{finite} hydration $\phi^*$ (Section~\ref{lam-stacks}) should be
simultaneously taken into account. This is not a too stringent
constraint since eq.~(\ref{eq:posm}) leads in the context of the
present model eq.~(\ref{eq:vdw1d-2d}) to a rather simple expression
for the osmotic pressure, namely
\begin{equation}
  \Pi=\frac{3\pi^2}{64}\frac{(k_BT)^2}{\kappa}\frac{1}{(\ell-\delta)^3}-k_BT\chi\phi^2
  \label{eq:posm1d-2d}
\end{equation}
where the terms originating from the two-dimensional fluid of lipid
molecules exactly cancel out. The resulting lamellar stack equation of
state, as well as the bilayer equation of state are displayed in
Fig.~\ref{fig:pi} and \ref{fig:sigma}, respectively.
The broad features of both figures are rather similar to the
experimental results displayed in Fig.~\ref{fig:parsegianPi} and
\ref{fig:parsegianSigma} for lamellar stacks, and (as far as the
pressure equation of state is concerned) in Fig.~\ref{fig:ethane} for
a simple fluid. In addition, the unbinding transition remains well
described within the framework of eq.~(\ref{eq:vdw1d-2d}), extending
Ref.~\cite{milner1992}. We have (numerically) explored the
consequences on the dilution limit $\ell_{\mathrm{max}}$ of decreasing
the Milner-Roux virial coefficient $\chi$, keeping all other
parameters fixed to the values given in Table~\ref{table:params}. As
shown in Fig.~\ref{fig:unbinding}, the prediction of our model is
compatible with a scaling law
$\ell_{\mathrm{max}}\propto|\chi-\chi_c|^{-\alpha}$, with critical
unbinding occurring for $\chi_c\approx2.2\times10^{-6}$~nm$^{-3}$ (or
$k_BT\chi_c\approx9.2$~Pa) and an unbinding exponent
$\alpha\approx0.9$.
The discrepancy between our result for the unbinding exponent,  namely
$\alpha\approx0.9$, and the theoretically-predicted values $\psi=1$ in
Ref.~\cite{milner1992} or $\psi=1.00\pm0.03$ in
Ref.~\cite{lipowski1986} may come from the unsophisticated numerical
methods used for obtaining the bilayer equation of state from
eq.~(\ref{eq:optSigma2}). We leave a more detailed discussion to
interested specialists.
\par In Section~\ref{exp_res}, we further strengthen the relevance of
our thermodynamic model, confronted with experimental results obtained
for lecithin--Simulsol lamellar phases of varying bilayer compositions.
\section{Experimental results and discussion}
\label{exp_res}
\begin{figure}
\begin{centering}
  \includegraphics[scale=1,width=\columnwidth]{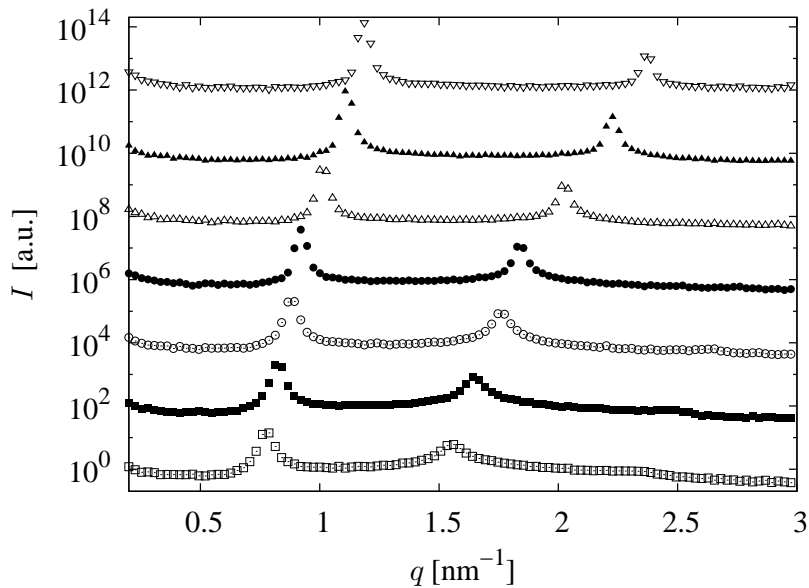}
\end{centering}
\caption{Small-angle x-ray scattering data for a series of lamellar
  stacks with bilayer composition $x=0.3$ in osmotic equilibrium with
  polymer solutions of increasing concentrations: $c_p=0.068$~g/mL
  ($\boxdot$), 0.100 ($\blacksquare$), 0.141 ($\odot$), 0.162
  ($\bullet$), 0.259 ($\triangle$), 0.378
  ($\blacktriangle$), and 0.472 ($\triangledown$). The stacking period
  $\ell$ decreases with increasing osmotic
  pressure \label{fig:saxsdata}}
\end{figure}

Small-angle x-ray scattering data are recorded, for a given bilayer
composition $x$ (described, from now on, in terms of Simulsol
\emph{mass} fraction, instead of mole fraction as in
Section~\ref{intro}), after immersing the dry lipid material in
polymer solutions of various initial concentrations and waiting for
water activity to equilibrate in both lamellar and polymer
phases. Typical results are displayed in Fig.~\ref{fig:saxsdata}, for
a system with Simulsol mass fraction $x=0.3$.
The first and second order Bragg peaks, with ratio 1:2 as expected for
a one-dimensional lamellar order, are always observed, a third order
peak being also noticeable in the investigated $q$-range for the more
hydrated samples. The Bragg peaks are shifted towards larger values as
osmotic pressure is increased because of the concomitant dehydration
of the lamellar stack.
\par
\begin{figure}
\begin{centering}
  \includegraphics[scale=1,width=\columnwidth]{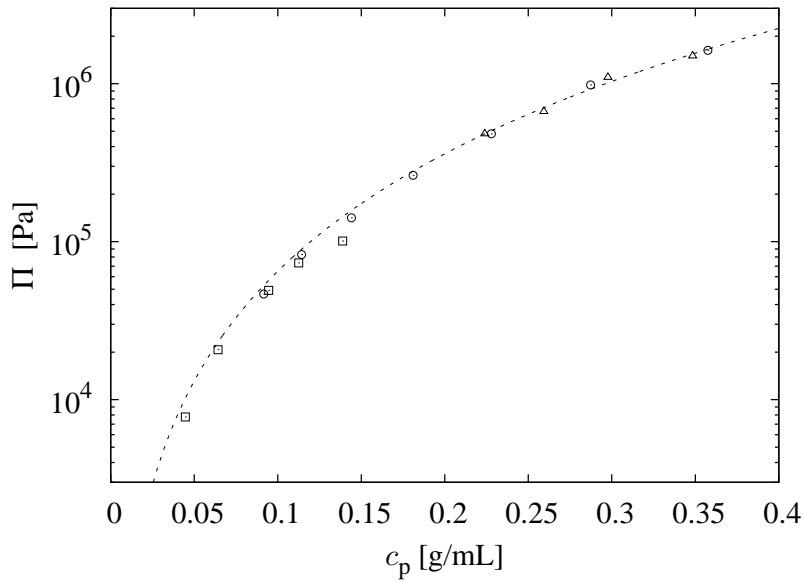}
\end{centering}
\caption{Osmotic pressure as a function of polymer concentration for
  three series of polymer solutions differing by the polymer degrees
  of polymerisation. PVP10, $\triangle$ -- PVP40, $\odot$ -- PVP360,
  $\boxdot$. The dotted line is a polynomial fit, appropriate for
  semi-dilute polymer solutions, with equation $\Pi=4\cdot10^6\times
  c_p^2\times(1+6.25\times c_p)$ -- pressure in Pa, concentration in
  g/mL
  \label{fig:piPol}}
\end{figure}

\begin{figure}
\begin{centering}
  \includegraphics[scale=1,width=\columnwidth]{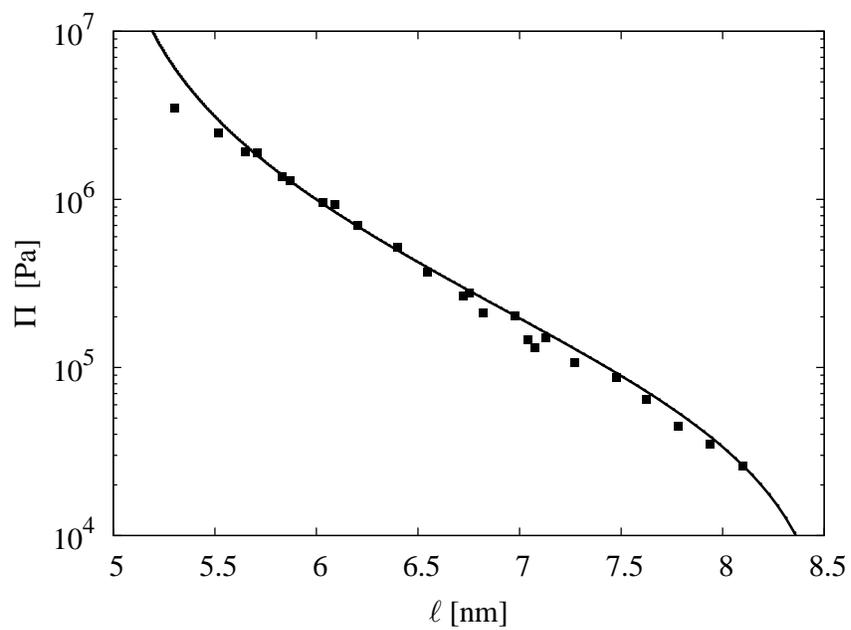}
\end{centering}
\caption{Lamellar stack equation of state $\Pi(\ell)$ for the system
  with bilayer composition $x=0.3$. Continuous line from model
  discussed in Section~\ref{model}, see text for details
  \label{fig:pi70-30}}
\end{figure}

 From the first order Bragg peak location $q_0$, the stacking
period $\ell$ is directly obtained: $\ell=2\pi/q_0$. Since the
dilution law, \emph{i.e.} the $\ell(\phi)$ relation, has been
previously experimentally determined~\cite{gerbelli2013}, the x-ray
measurement indirectly gives the bilayer volume fraction
$\phi$. Because the amount of (initially dry) lipid in the whole
system is known, mass conservation allows determining the
\emph{equilibrium} polymer concentration from its initial value in the
known amount of polymer solution used for hydrating the lamellar
stack. The lamellar stack equation of state then results from the
auxiliary calibration curve (see Fig.~\ref{fig:piPol}) giving the
osmotic pressure $\Pi$ of the polymer solution as a function of
polymer concentration.
An illustration for the system with Simulsol mass fraction $x=0.3$ is
given in Fig.~\ref{fig:pi70-30}, with a quasi-exponential decay of the
osmotic pressure observed in a range of stacking parameters $\ell$.
\par
\begin{figure}
\begin{centering}
  \includegraphics[scale=1,width=\columnwidth]{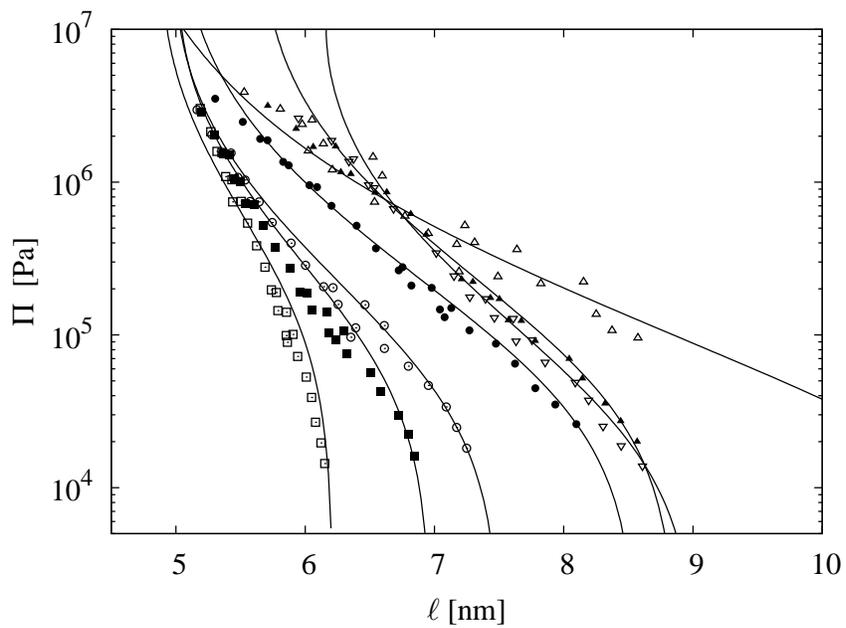}
\end{centering}
\caption{Lamellar stack equations of state $\Pi(\ell)$ for systems
  with bilayer compositions $x=0$ ($\boxdot$), $x=0.05$
  ($\blacksquare$), $x=0.1$ ($\odot$), $x=0.3$ ($\bullet$), $x=0.5$
  ($\triangle$), $x=0.7$ ($\blacktriangle$) and $x=0.8$
  ($\triangledown$). Continuous lines from model discussed in
  Section~\ref{model}, see text for details
  \label{fig:piAll}}
\end{figure}

 The set of data shown in Fig.~\ref{fig:piAll} is obtained by
repeating the same procedure and analyses for mass fractions spanning
the pure lecithin ($x=0$) to the almost pure Simulsol ($x=0.8$) range.
It was not possible to include the \emph{pure} Simulsol system, and
not even the $x=0.9$ bilayer composition in the present study because
at low osmotic pressures the water channel height reaches values so
large that the polymer molecules can no longer be considered as
\emph{excluded} from the lamellar stack. Thus, increasing $x$
apparently plays the r\^ole of approaching the unbinding transition
here, as recently observed for a closely related system by a different
technique~\cite{bougis2015}.
\par
\begin{figure}
\begin{centering}
  \includegraphics[scale=1,width=\columnwidth]{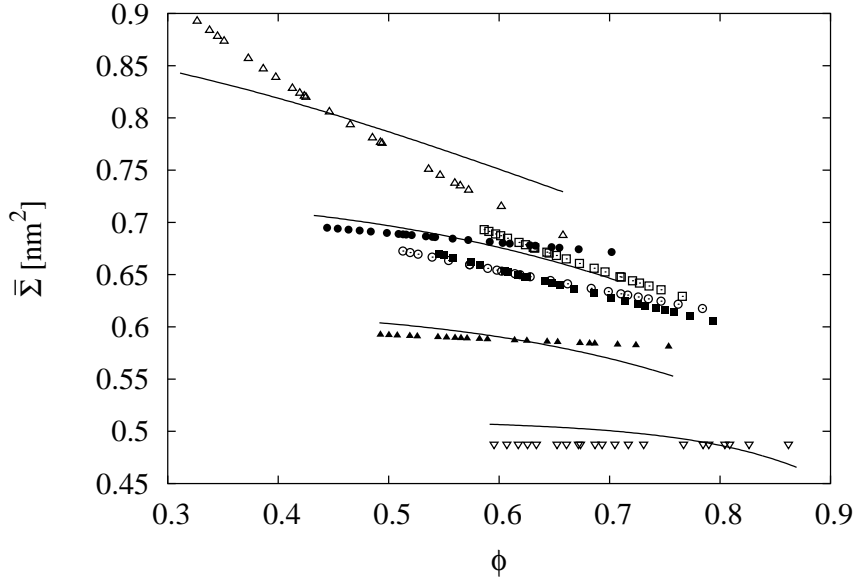}
\end{centering}
\caption{Experimental (symbols) and selected model (continuous lines)
  bilayer equations of state $\bar{\Sigma}(\phi)$ for lamellar stacks
  with bilayer compositions $x=0$ ($\boxdot$), $x=0.05$
  ($\blacksquare$), $x=0.1$ ($\odot$), $x=0.3$ ($\bullet$, ---),
  $x=0.5$ ($\triangle$, ---), $x=0.7$ ($\blacktriangle$, ---) and
  $x=0.8$ ($\triangledown$, ---)
  \label{fig:sigmaAll}}
\end{figure}

 The thermodynamic model discussed in Section~\ref{model} and,
more specifically, eq.~(\ref{eq:vdw1d-2d}) have been used for
describing the lamellar stack equations of state displayed in
Fig.~\ref{fig:pi70-30} and \ref{fig:piAll}, as well as the bilayer
equations of state displayed in Fig.~\ref{fig:sigmaAll}.
We have not attempted to implement a complete fitting procedure to
finely tune values for \emph{all} the parameters listed in
Table~\ref{table:params}, but checked in a few numerical simulations
the influence of changing the bilayer bending modulus $\kappa$
(bilayer stack entropy), as well as the Milner-Roux virial coefficient
$\chi$ (bilayer stack enthalpy, resulting from both attractive and
repulsive contributions to inter-bilayer \emph{direct} interactions)
and excluded area $\Sigma_0$ (two-dimensional fluid entropy). The
continuous curves superimposed to data points in
Fig.~\ref{fig:pi70-30} and \ref{fig:piAll} obviously give a fair
description of the osmotic pressure data, with parameters given in
Table~\ref{table:exp}.
\begin{table}
  \centering
  \begin{tabular}{|c||c|c|c||c|} \hline\hline
$x$ & $\kappa/k_BT$ & $\chi/\varsigma^3$ & $\Sigma_0$ &
$\ell_{\mathrm{max}}$ \\ \hline
    &               & [nm$^{-6}$]         & [nm$^2$]   & [nm] \\ \hline\hline
0.8 & 0.55          & 0.18               & 0.370      & 9.01 \\ \hline
0.7 & 0.15          & 0.43               & 0.477      & 8.86 \\ \hline
0.5 & 0.047         & 0.43               & 0.635      & 12.0 \\ \hline
0.3 & 0.13          & 0.43               & 0.525      & 8.53 \\ \hline
0.1 & 0.33          & 0.30               & 0.495      & 7.51 \\ \hline
0.05 & 0.40         & 0.36               & 0.485      & 6.95 \\ \hline
0.0 & 0.35          & 0.65               & 0.505      & 6.22 \\ \hline
  \end{tabular}
  \caption{Numerical values chosen for describing experimental
    lamellar stack equations of state using
    eq.~(\ref{eq:vdw1d-2d}). Parameters $\bar{v}$ and $b_2/\Sigma_0$
    kept constant to values 1.25~nm$^3$ and $-10$,
    respectively. Parameter $\varsigma\equiv\sqrt{\Sigma_0}$. The last
  column gives the \emph{model-predicted} dilution limits
  $\ell_{\mathrm{max}}$ that result from the chosen parameters}
  \label{table:exp}
\end{table}
The description, though acceptable, is less satisfactory as far as the
bilayer equation of state is concerned, see Fig.~\ref{fig:sigmaAll}.
Note that, for a better readability, a (representative) subset only of
the available model curves is displayed. Improvements in the
two-dimensional liquid model for describing the intra-bilayer
thermodynamic contribution to the lamellar stack excess free energy,
eq.~(\ref{eq:vdw1d-2d}), might here be required.
\par
\begin{figure}[h]
  \centering \subfloat[][Bending modulus
  $\kappa$]{\includegraphics[scale=1,width=0.45\columnwidth]{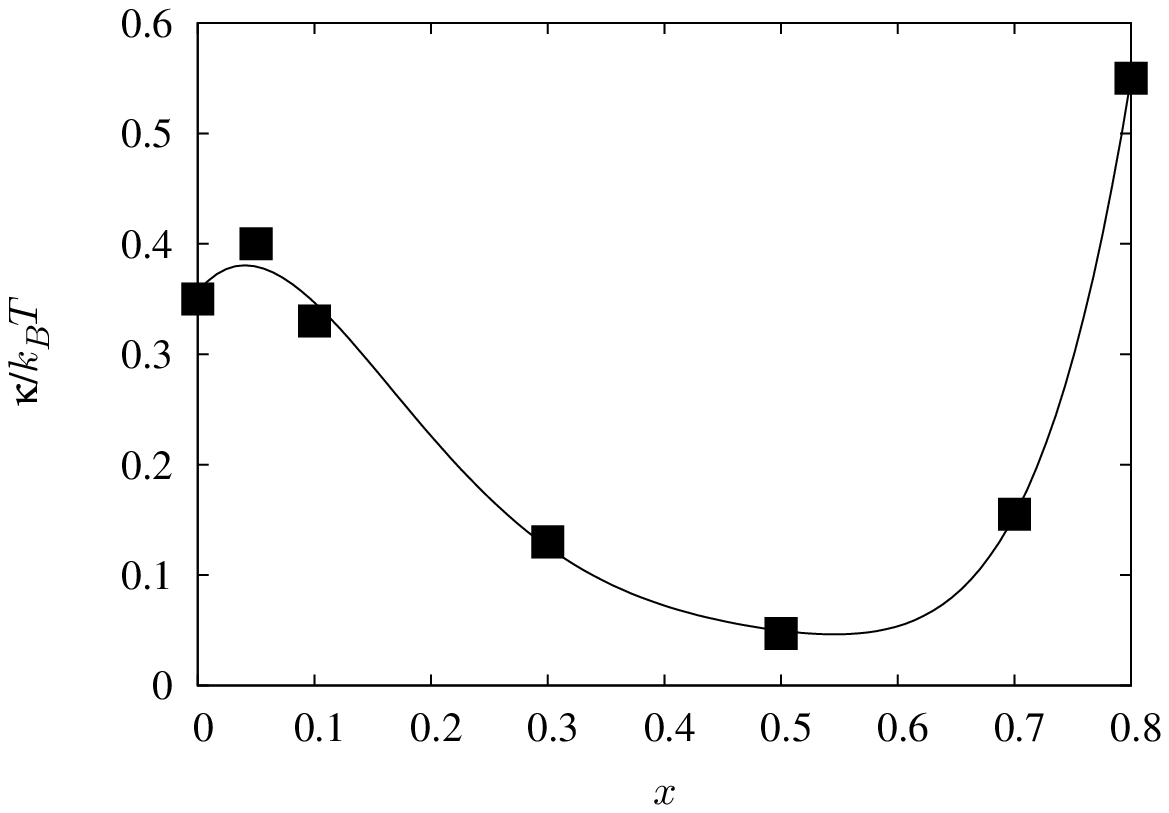}\label{fig:params_kappa}}
  \qquad \subfloat[][Milner-Roux second virial coefficient
  $\chi$]{\includegraphics[scale=1,width=0.45\columnwidth]{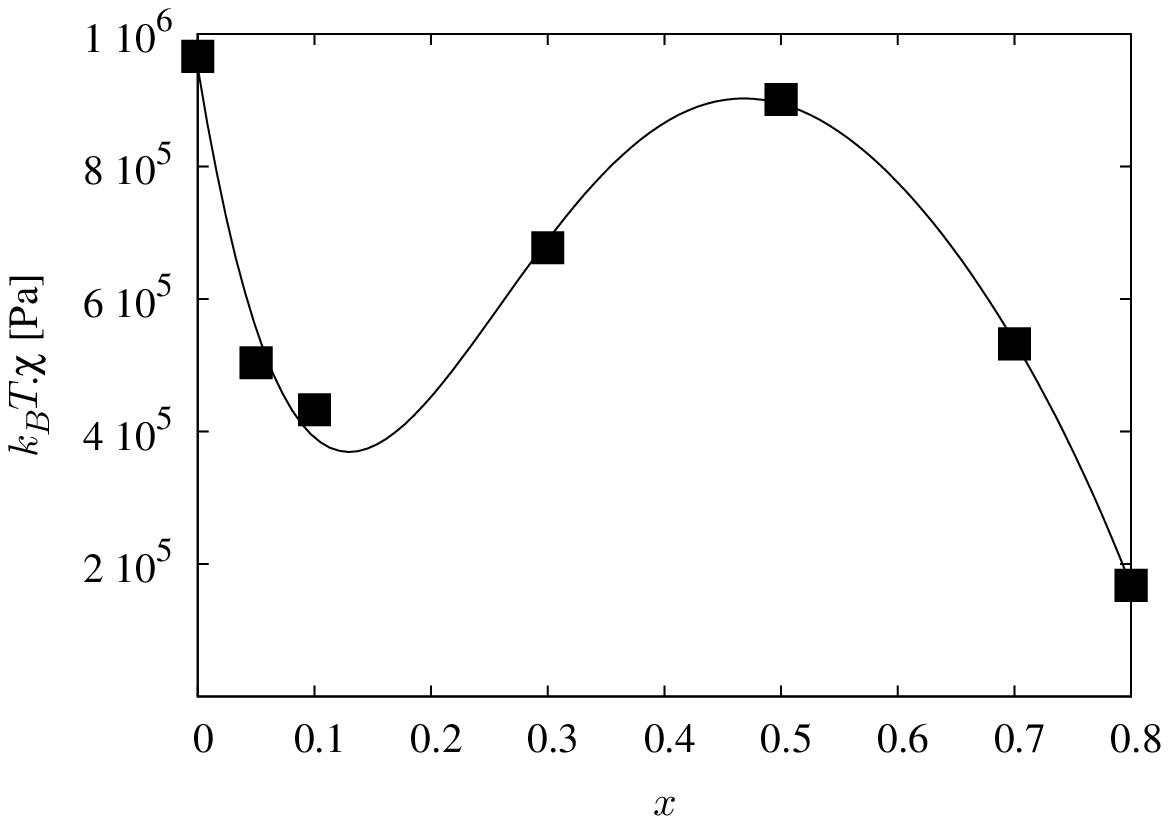}\label{fig:params_chi}}
  \\ \subfloat[][{Excluded area
    $\Sigma_0$}]{\includegraphics[scale=1,width=0.45\columnwidth]{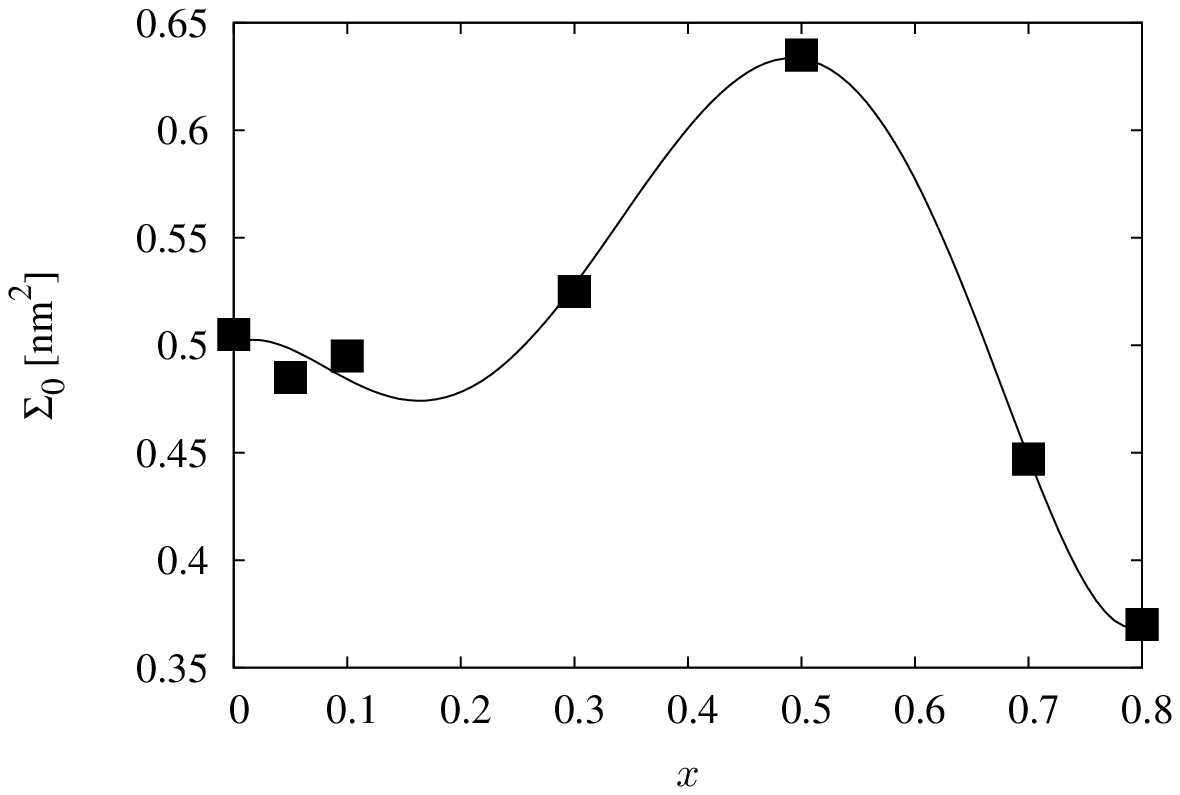}\label{fig:params_sigma0}}
  \qquad \subfloat[][Dilution limit
  $\ell_{\mathrm{max}}$]{\includegraphics[scale=1,width=0.45\columnwidth]{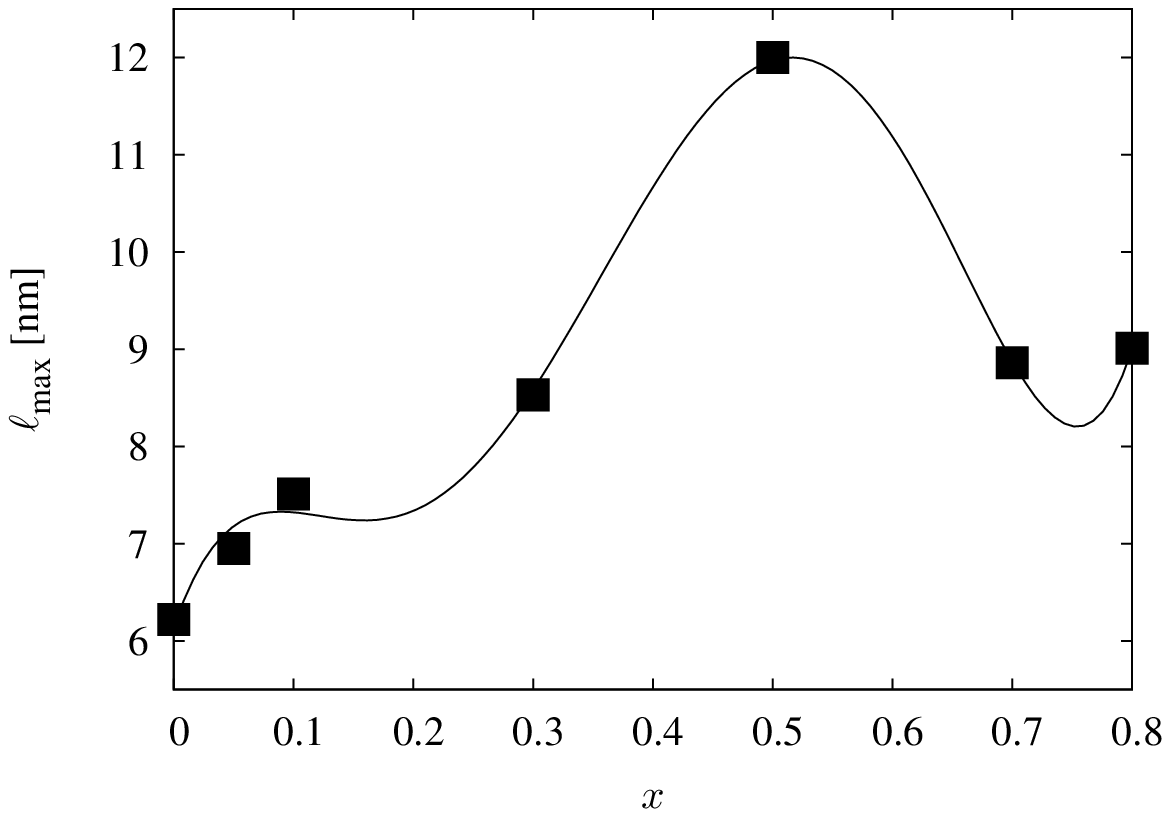}\label{fig:params_lmax}}
  \caption{Evolution as a function of bilayer composition $x$ of the
    lamellar stack thermodynamic parameters. The continuous lines are
    guides for the eye. The bilayer configurational entropy is
    maximal close to
    $x=0.5$--Fig.~\ref{fig:params_kappa}. Inter-bilayer attractive
    interactions are maximal and of comparable magnitude for $x=0$ and
    $x=0.5$--Fig.~\ref{fig:params_chi}. The excluded area of the
    two-dimensional bilayer fluid is maximal close to
    $x=0.5$--Fig.~\ref{fig:params_sigma0}. The dilution limit of the
    lamellar stack is maximal close to
    $x=0.5$--Fig.~\ref{fig:params_lmax}}
\label{fig:params}
\end{figure}

 There is a rather remarkable trend in the evolution with the
bilayer composition $x$ of the model parameters $\kappa$, $\chi$ and
$\Sigma_0$, as well as in the model-predicted dilution limit
$\ell_{\mathrm{max}}$ illustrated in Fig.~\ref{fig:params_kappa},
\ref{fig:params_chi}, \ref{fig:params_sigma0}, and
\ref{fig:params_lmax} respectively.
Two conspicuous compositions can be distinguished, one in the range
$x=0.05$--$0.1$ close to \emph{pure} lecithin, and when $x$ is close
to 0.5. In a previous structural study of the same lecithin--Simulsol
system~\cite{gerbelli2013}, the value $x=0.5$ was already
experimentally evidenced as separating two regimes, even though the
\emph{experimental} dilution limit did not follow what is predicted
here Fig.~\ref{fig:params_lmax}. Still, the data for the area per
(effective) lipid molecule $\bar{\Sigma}$ as a function of hydration
displayed in Fig.~\ref{fig:sigmaAll} clearly indicates that
composition $x=0.5$ experimentally plays a special r\^ole. The area
per lipid molecule is significantly larger--and specially sensitive to
hydration--than with all other bilayer compositions, an experimental
feature independent from any model. This could indicate that the
composition threshold for the ``brush-to-mushroom'' conformational
transition in the hydrophilic blocks of Simulsol molecules suggested
in Ref.~\cite{gerbelli2013} and \cite{bougis2015} is close to $x=0.5$,
and may give a clue for reaching a better agreement between
experimental and predicted bilayer equations of state, at the expense
of introducing a ``brush-to-mushroom'' variable into
eq.~(\ref{eq:vdw1d-2d}). Besides, the experimental ``back-and-forth''
variation of $\bar{\Sigma}$ when $x$ is close to zero observed in
Fig.~\ref{fig:sigmaAll}, though not easily interpreted, is nicely
compatible with the picture that emerges from the thermodynamic model
of the lamellar stack, Fig.~\ref{fig:params}, being simultaneously
much less model-dependent: It should therefore correspond to a robust,
but still unexplained, feature of the lecithin--dilute Simulsol
mixture.
\section{Conclusion}
\label{conclusion}
The interplay between soft confinement and steric effects has been
studied in lamellar stacks of mixed lecithin--Simulsol bilayers,
varying hydration and bilayer composition, by coupling osmotic
pressure with small-angle x-ray scattering measurements that give
access to both ``lamellar stack'' and ``bilayer'' equations of
states. 
\par With the help of a thermodynamic model of lamellar stacks that
extends the approach of S.T. Milner and D. Roux~\cite{milner1992} by
the explicit inclusion of the thermodynamic properties of a
two-dimensional fluid representing the lipid bilayers, we are able to
quantitatively reproduce the osmotic pressure, and semi-quantitatively
the lipid area data, keeping a meaningful description of the unbinding
transition. The physical ingredients of the model are the Helfrich
undulation interactions, controlled by a bending modulus parameter
$\kappa$, \emph{direct} inter-bilayer interactions controlled by the
Milner-Roux virial coefficient $\chi$, and a description of the
two-dimensional lipid fluid by the venerable van der Waals model
designed more than a century ago for ordinary fluids, with an excluded
area parameter $\Sigma_0$ and a second virial coefficient $b_2$. The
celebrated, but elusive ``hydration forces'' play no r\^ole whatsoever
in the model, while quasi-exponential osmotic pressure profiles
remain, at least in a restricted range of water content in the
lamellar stacks.
\par Among the intriguing outcomes of applying the model to describing
the available osmotic and small-angle scattering data, it appears that
adding a co-surfactant (Simulsol) to the lipid bilayer \emph{not
  necessarily} always increases the bilayer flexibility, even though
$\kappa$ remains very sensitive to the amount of co-surfactant. This
may be due to a kind of ``brush-to-mushroom'' transition taking place
in the hydrophilic block of the Simulsol surfactant. The model
prediction also does not quite fit with previous experimental results
on the same system as far as the \emph{dilution limit} of the lamellar
stacks are concerned.
\par The better knowledge of the lamellar stack free energy gained
here should still be refined, for instance in view of understanding
the detailed mechanisms of the lamellar-lamellar phase coexistence at
low hydration observed in rather similar lecithin--Simulsol
systems~\cite{bougis2015}. 
\section*{Acknowledgements}
The support by Funda\c c\~ao de Amparo \`a Pesquisa do Estado de S\~ao
Paulo through grant 2011/16149-8 is gratefully acknowledged.  We also
thank IdEx Bordeaux (France) and Conselho Nacional de Desenvolvimento
Cient\'\i fico e Tecnol\'ogico (Brazil) for providing support through
their respective programs ``Doctorat international'' and ``Ci\^encia
sem Fronteiras''.
\section*{Author contribution statement}
%
All authors contributed equally to the present work.
\appendix
\section{Appendix}
\label{appendix}
\subsection{Polymer solutions}
\label{PolSol}
The polymer used in this study is polyvinylpyrrolidone (PVP), from
Sigma-Aldrich, in three molecular weights (10, 40 and 360~kg/mol). The
polymer solutions are purified using snakeskin dialysis tubes (from
Thermo Scientific) of different pore dimensions. The dialysis
compartments filled with polymer solution were immersed in a tank
initially containing pure water, and the water content was changed a
few times, until the total volume reached 300 times the volume of the
solution. The whole procedure took about one week. Once purified, the
solutions were freeze-dried, which allowed re-dispersing the polymer
material in water at various desired concentrations. The complete
polymer dissolution could take a few days for the more concentrated
solutions with PVP360.
\par The osmotic pressure of each prepared solution was measured with
the PZL-100 osmometer from PZL Tecnologia Company. Results are
displayed in Fig.~\ref{fig:piPol}. The nearly-quadratic increase of
osmotic pressure with polymer concentration, independently of the
polymer molar mass, is characteristic of polymer solutions being in
their semi-dilute regime. The $\Pi(c_p)$ relation is empirically
described by the polynomial law $\Pi=4\cdot10^6\times
c_p^2\times(1+6.25\times c_p)$ -- pressure in Pa, concentration in
g/mL -- which allows computing the osmotic pressure applied to
lamellar stacks.
%
%
\subsection{Osmotically-stressed lamellar stacks}
\label{BilPre}
The lipid membranes are prepared with soy lecithin (Avanti Polar
Lipids) and a non-ionic commercial cosurfactant (Simulsol 2599 PHA,
Seppic). Soy lecithin contains mainly dilinoleoylphosphatidylcholine
(DLPC), with typically 35\% of other (zwitterionic) lipids. Simulsol
is a mixture of ethoxylated fatty acids (71\% oleic and 11\% palmitic
acids being the main components), with an average of 10
CH$_2$--O--CH$_2$ groups in the hydrophilic block. Lecithin and
Simulsol are co-solubilised in cyclohexane in desired proportions,
which always leads to macroscopically homogeneous, transparent
  solutions. Therefore, \emph{a priori} homogeneous
mixtures of controlled composition (labelled as a mass fraction $x$)
are obtained by evaporating the solvent. Lamellar systems are then
prepared by swelling a chosen mass of the dry amphiphilic mixture with
thrice the amount of polymer solutions, tuning the polymer mass and
initial concentration to scan as much as possible of the accessible
domain of osmotic pressures once equilibrium is reached.  The sample
tubes were conserved at $4\,^{\circ}$C and cycles of centrifugation
were done to accelerate homogenisation. Equilibrium is reached when
water activity is the same in coexisting lamellar and polymer phases,
which is easily deduced by observing the tubes characterised by a
viscous and slightly turbid solution on top of a transparent polymer
solution. Osmotic equilibrium also implies that \emph{some}
  lecithin or Simulsol species are to be found in the polymer
  solution. The corresponding concentrations are, however, expected to
  be in the order of their respective critical
  micellar--aggregation--concentrations, viz. quite low. Neglecting
  the amount of ``lost'' material in computing the lamellar phase
  lipid fraction, as well as neglecting the contribution of lipid
  aggregates to osmotic pressure are therefore safe approximations.
\subsection{Small-angle x-ray scattering}
\label{xray}
The small-angle scattering experiments were carried out at the
Institute of Physics, University of S\~ao Paulo, Brazil, with the
Xeuss instrument form Xenocs equipped with a Pilatus 300K detector
(Dectris).  Radiation produced by the microfocus Copper source is
collected with a single-reflection multilayer optic producing a
low-divergence, monochromatic beam with wavelength $\lambda=0.154$~nm
that is further collimated by a pair of scatterless slits--upstream
slits 0.6$\times$0.6~mm, downstream slits 0.5$\times$0.5~mm. The
sample-to-detector distance, calibrated with a Silver Behenate
standard, is 0.77~m. Detector images span in practice a scattering
wave vector range extending from 0.04~nm$^{-1}$ to
3.5~nm$^{-1}$
. Samples are held in glass
capillaries with a nominal diameter of 1.5~mm, and the scattering
intensity is corrected for background by subtracting the properly
normalised signal of a capillary filled with pure water. Exposure time
varies from 15 up to 30 minutes, depending on the polymer solution
concentration.
\end{document}